%% file: coling_latex.tex
\pdfoutput=1

\documentclass[11pt]{article}
\DeclareUnicodeCharacter{202A}{!!!FIX ME!!!} 
\usepackage[]{coling}
\usepackage{times}
\usepackage{latexsym}
\usepackage{booktabs}

\usepackage[T1]{fontenc}

\usepackage[utf8]{inputenc}

\usepackage{microtype}

\usepackage{inconsolata}

\usepackage{graphicx}
\usepackage{amsmath}
\title{AmalREC: A Dataset for Relation Extraction and Classification Leveraging Amalgamation of Large Language Models}

\author{
  Mansi, Pranshu Pandya, Mahek Bhavesh Vora, Soumya Bharadwaj, Ashish Anand \\
  Department of Computer Science and Engineering\\
  Indian Institute of Technology, Guwahati \\
    \texttt{\{m.mansi, p.pandya, v.mahek\}@alumni.iitg.ac.in} \\
\texttt{\{soumya.bharadwaj, anand.ashish\}@iitg.ac.in}, \\
  \small{\textbf{Correspondence:} \href{mailto:soumya.bharadwaj@iitg.ac.in}{soumya.bharadwaj@iitg.ac.in}}
}

\begin{document}
\maketitle
\begin{abstract}
Existing datasets for relation classification and extraction often exhibit limitations such as restricted relation types and domain-specific biases. This work presents a generic framework to generate well-structured sentences from given tuples with the help of Large Language Models (LLMs). This study has focused on the following major questions: (i) how to generate sentences from relation tuples, (ii) how to compare and rank them, (iii) can we combine strengths of individual methods and amalgamate them to generate an even better quality of sentences, and (iv) how to evaluate the final dataset? For the first question, we employ a multifaceted \textbf{5-stage pipeline} approach,
leveraging LLMs in conjunction with template-guided generation. We introduce \textbf{Sentence Evaluation Index(SEI)} that prioritizes factors like grammatical correctness,
fluency, human-aligned sentiment, accuracy, and complexity to answer the first part of the second question. To answer the second part of the second question, this work introduces a \textbf{SEI-Ranker module} that leverages SEI to select top
candidate generations. The top sentences are then strategically amalgamated to produce the final, high-quality sentence. Finally, we evaluate our dataset
on LLM-based and SOTA baselines for relation classification. The proposed dataset features \textbf{255 relation types}, with 15K sentences in the test set and around 150k in the train set organized in, significantly enhancing relational diversity
and complexity. This work not only presents a new comprehensive benchmark dataset for RE/RC task, but also compare different LLMs for generation of quality sentences from relational tuples.
\end{abstract}

\section{Introduction}

Relation extraction and classification (RE/RC) is important for natural language understanding (NLU), underpinning tasks like information extraction (IE), knowledge base construction (KBC), and question answering. Relation Extraction is the task of identifying possible entities and relations between the entities, whereas Relation Classification is the task of classifying the relation between the given entities from a set of pre-defined relations \citep{hendrickx2019semeval}. 
 The quality of RE/RC directly impacts the effectiveness of downstream applications. Moreover, knowledge bases (KBs) and information retrieval (IR) techniques play a pivotal role in retrieval-augmented generation (RAG) models. By improving RE/RC, we empower RAG systems to better understand and leverage relevant knowledge, leading to more informed and accurate responses \citep{lewis2020retrieval}.

However, the effectiveness of the RE/RC models heavily relies on the quality and diversity of the datasets used for training. The existing relation classification datasets, such as TACRED \citep{zhang2017tacred}, CoNLL04 \citep{roth2004linear}, and NYT \citep{zhu2020towards} have a restricted number of relation types, hindering models' ability to capture the diversity of relationships found in real-world language. Additionally, datasets like CoNLL04 and SciREC \citep{luan2018multi} may have domain-specific biases.  Moreover, datasets like DocRED \citep{yao2019docred}, while constructed using distant supervision, can sometimes exhibit limitations in the completeness and consistency of their annotations. 

To address these shortcomings, we have created a novel dataset for relation classification that significantly expands the scope of existing resources. Recognizing the need to challenge current state-of-the-art models, our dataset increases both the number of relations and the complexity of the relationships represented. We incorporate a diverse set of 255 unique relation types derived from DBpedia \citep{dbpedia-swj} by \citep{parekh2019taxonomical}. 
Our dataset \texttt{AmalREC} has been generated by amalgamating results across 15 methods encompassing finetuning-based, prompt-based, template-based, fusion-based, and extended context-based (ECB) generation paradigms. We have explained the generation methods in detail in section ~\ref{sec:method}. For each tuple, we generate sentences using these 15 methods and rank the sentences using proposed Sentence Evaluation Index (SEI). SEI takes into account grammatical correctness, readability, human-aligned sentences, fluency, relevance, accuracy, and logical coherence. We take the top three sentences ranked by the SEI-ranker and amalgamate the top 3 sentences along with the gold and ECB-generated sentences to create the final dataset.

Finally, we evaluate our dataset on Relation extraction and classification tasks on state-of-the-art methods like TANL \citep{simi-attardi-2016-adapting} and SPERT \citep{eberts2019span} along with leading open-source and proprietary LLM based classification techniques inspired by \citep{xu2023unleash} and \citep{wadhwa2023revisiting}, leveraging LLMs like Mixtral, Palm and GPT3.5. We further perform human verification on a subset of the test set, and perform ablation studies to study and justify the need for each stage in our pipeline. 

In summary, we make the following contributions:
\begin{itemize}
    \item \textbf{Dataset AmalREC:} We present a relation extraction/classification dataset with more relations, generated from the best of all the models/ generation techniques. We also provide baseline results for relation classification task on SOTA models.
    \item \textbf{Comparison of LLMs: } As this study uses several generation methods using LLMs, a comparative analysis is performed to obtain top performing generation models. 
    \item \textbf{Quality-score-based ranking and blending technique} to generate the best sentence from existing generations.
\end{itemize}

\section{Related Work}
\textbf{Generation techniques: }As the primary aim is to generate a sentence from a given relation tuple, we first discuss the relevant works on text generation from a tuple. \citep{reiter_dale_2000} proposed a \textbf{rule-based method} which decomposes the task into a five-stage sequential pipeline relying heavily on trained linguistics to develop rules and were replaced by templates and other neural methods.

Later came end-to-end neural methods like \citep{castro-ferreira-etal-2019-neural, dusek-e2e} which used Resource Description Format (RDF), graphs, and tables as inputs. Such methods utilize diverse implicit structures, such as content planning in \citep{Puduppully19} and linear structure of graphs encoded in \citep{zhao-etal-2020-bridging} to generate sentences. \citep{laha-etal-2019-scalable} presented a semi-template-based approach. \citep{harkous-etal-2020-text} proposed an automatic generation approach from structured data ensuring semantic fidelity.

\citep{kale-rastogi-2020-text} was the first to introduce T5 for the task. \citep{xu-etal-2023-unleash} further strengthened the usage of LLMs RE and RC tasks. \citep{wadhwa2023revisiting} investigated context learning for few shot relation extraction utilizing GPT 3/3.5 and Flan T5, and \citep{ribeiro-etal-2021-investigating} compared efficacy of neural network-based methodologies against conventional computational pipelines. Studies in \citep{li-etal-2023-revisiting-large, ma-etal-2023-chain-thought} incorporated chain of thought prompting techniques to improve the zero/few-shot RE.

This research introduces a novel multifaceted approach that amalgamates strengths of LLMs and other approaches for RE/RC dataset generation inspired by \citep{jiang2023llmblender, kasner-dusek-2020-data}. 

\textbf{RE/RC Datasets:} Datasets like \textbf{TACRED} \citep{zhang2017tacred} employ automated methods in conjunction with manual supervision, while others  \textbf{DocRED} \citep{yao2019docred}, \textbf{RedFM} \citep{redfm2023} create their systems using data from Wikipedia and Wikidata. \textbf{NYT} \citep{zhu2020towards} uses distant supervision method.
The majority of these datasets cover a limited range of relation types and have no control over the quality of generations. Our dataset tackles these issues by providing a diverse set of real-world relations and quality sentences based on a rich set of quality parameters, challenging the existing SOTA RE/RC baselines. 

\section{Problem Formulation and Methodology}
\subsection{Problem Statement}
Our goal is to generate a dataset for relation extraction and classification, $D = \{xi | \forall i \}$. Each instance in the dataset is represented as a tuple consisting of a sentence and a relation tuple: $x_i = (s_i, rel_i)$. Here, $s_i$ is the sentence, and $rel_i$ is the relation tuple identified within that sentence.

The relation tuple $rel_i$ is defined as $rel_i$ = ($e1_{i}$, $t1_{i}$, $r_i$, $e2_{i}$, $t2_{i}$), where:
\begin{align*}
    s_i &: \text{i}^{th} \text{ sentence,}\\
    e1_{i} &: \text{First entity in } s_{i},\\
    t1_{i} &: \text{Type of } e1_{i},\\
    e2_{i} &: \text{Second entity in } s_{i},\\
    t2_{i} &: \text{Type of } e2_{i},\\
    r_{i} &: \text{Relation between } e1_{i} \text{ and } e2_{i}.
\end{align*}

A relation bucket $rb_k$ is a collection of relation tuples $rel_i$ where the corresponding entities $(e1_i, e2_i)$ share the same tuple $(t_{1k}, r_k, t_{2k})$. Formally, the relation bucket is defined as $rb_k = {(e_{1i}, t_{1k}, r_k, t_{2k}, e_{2i}) \forall i }$. In our dataset, we are considering only three named entity types, namely, Person, Location and Organization.

\subsection{Methodology}
\label{sec:method}
\begin{figure*}[t]
    \centering
    \includegraphics[height=0.50\textheight]{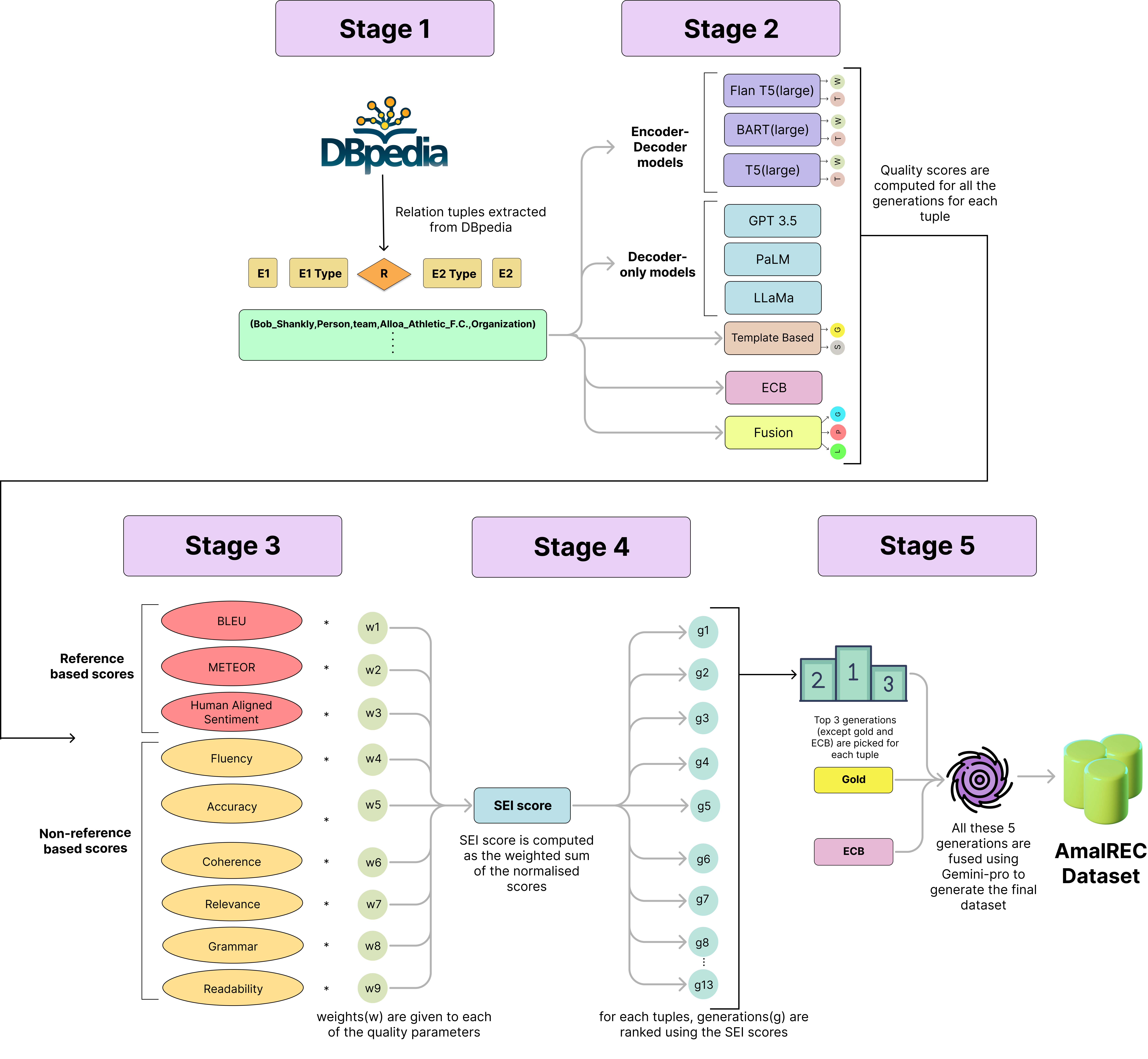} 
    \caption{The 5-stage pipeline; Abbreviations in stage 2, Encoder-Decoder based models: T: Tacred, W: WebNLG based fine-tuning; Template based generation: G: Gold, S: Silver; ECB: Extended context based generation; Fusion: G: GPT 3.5, P: PaLM, L: LLaMa datasets based fine-tuning on Flan T5}
    \label{fig:enter-label}
\end{figure*}

Our task is to generate a \textit{grammatically correct}, \textit{sentimentally human aligned}, \textit{readable}, \textit{fluent}, \textit{accurate}, \textit{relevant} and \textit{coherent} sentence from a given relation tuple. The proposed data generation pipeline is divided into 5 stages. Figure~\ref{fig:enter-label} illustrates the five stages. The subsequent sections now discuss each of these stages.
\subsubsection{\textbf{Stage 1: Tuple collection}}
In Stage 1, we start by the collection of relation tuples. The relation tuples are taken from the dataset shared by \citep{parekh2019taxonomical}, which has extracted the relation tuples from DBpedia. The dataset consists of 255 distinct relation buckets. The dataset spans more than 5.9 million instances. 


From this extensive dataset of relation tuples, we filter a balanced subset by relation bucket, selecting approximately 2,000 tuples per bucket to create an overall dataset of around 195,000 relation tuples.From this set, we segregate 15k instances spanning equally across all the 255 relation buckets for curating an inclusive \textit{test set} and, we use the rest for \textit{train} and \textit{dev} set generation. We will be using these 195k relation tuples for generation of our final dataset $D$.

\subsubsection{\textbf{Stage 2: Multifaceted data generation}}
\label{sec:sent_gen}
Now, after collecting a balanced dataset of relation tuples $rel_i$, we move on to Stage 2 where we focus on answering the major question of how to generate sentences from a given tuple. We utilize a novel multifaceted approach for data generation via 15 different methods. Broadly, we have generated sentences in 5 different paradigms, as explained below.

\textbf{Template based generation: }
\label{sec:temp_gen}
In this method, sentence templates are generated by humans for each of the 255 relation buckets to generate sentences for all the relation tuples in the bucket. These templates are generated such that filling in the instances from the bucket would produce new sentences. For example, a template for the relation bucket: \texttt{(Location, administrative district, Location)}, is ``\textless $e1_{i}$\textgreater is an administrative district in \textless $e2_{i}$\textgreater."

To generate the \textbf{Gold standard} dataset, we filled all instances for each relation bucket, but this resulted in similar sentence types within each bucket. To introduce syntactical diversity, we used two paraphrasing models, \textbf{Humarin} \citep{chatgpt_paraphraser} and \textbf{Pegasus} \citep{zhang2019pegasus}, paraphrasing half of the sentences with each. This approach added variation in syntactical structures, creating a more nuanced set of sentences while maintaining coherence with the original templates. These paraphrased sentences form the \textbf{``Silver standards"}.

\textbf{Encoder-decoder models based generation: }
This technique leverages three encoder-decoder models namely \textbf{T5(large)} \citep{raffel2023exploring}, \textbf{Flan T5(large)} \citep{chung2022scaling}, \textbf{BART(large)} \citep{lewis2019bart} for generating sentences from relation tuples. We prompt these models in Zero-shot and few-shot settings, using the \textbf{CoT prompting} technique. The results were unsatisfactory in terms of variations across generations, as instruction tuning these models is not always reliable. Therefore, we shifted to fine-tuning the models on \textbf{TACRED} (with a slight modification) \citep{zhang2017tacred} and \textbf{WebNLG} \citep{gardent2017webnlg} datasets. As TACRED is noisy \citep{stoica2021retacredaddressingshortcomingstacred}, we selectively chose a relatively clean subset, by taking sentences with at-least 7 and at-max 64 tokens. These sentences were then paraphrased using Pegasus \citep{zhang2019pegasus}. For WebNLG, we chose the 1-triple set. Finally, we finetune all the 3 models, separately on these two datasets to form two variations of each model. These finetuned models were then used to generate sentences for our relation tuples in a Few-shot setting. Appendix \ref{sec:prompts} discusses the prompts used for this method. Ablation study was conducted to establish firm necessity for finetuning, the details of which can be found in appendix \ref{sec: Stage 2 finetuning ablation}

\textbf{Decoder-only models based generation: }
In this generation technique, we use the decoder-only models, namely \textbf{GPT 3.5, LLaMa} \citep{touvron2023llama} and \textbf{PaLM} \citep{chowdhery2023palm}, to generate complex and diverse sentences given the relation tuples. We again use CoT~\citep{wei2022chain} prompting techniques in a few-shot setting. In the case of GPT 3.5, we generated the results in batches to reduce cost, and promote better quality generation. We notice that as the generated content by LLMs became longer, the quality decreased, in terms of complexity of the sentence as well as accuracy of representing the relations between the entities, also observed in \citep{li2023semiautomaticdataenhancementdocumentlevel}. To mitigate this, we empirically observe that the batch size of 60 was a good trade-off. However, we still faced issues such as incorrect mapping of sentences to tuples and merging of sentences. To mitigate this concern, we devise a mapping module for getting the correct mapping, details of which can be found in appendix \ref{sec: mapping module}.


\textbf{Fusion Technique:}
We observe that the encoder-decoder-models, like Flan T5, tend to generate more concise and simple sentences. In contrast, the Decoder-only models, like GPT 3.5, tend to generate more wordy and complex sentences. We employ a fusion technique to exploit the advantages of both models. 
The fusion technique fine-tunes Flan T5 on sentences generated by decoder-only models (GPT-3.5, LLaMa, and PaLM) from a secluded set of triples, carefully chosen to avoid overlap with the test set. The fine-tuned model is then used to generate sentences for new test-set tuples. An ablation study, detailed in Appendix \ref{sec: Stage 2 fusion ablation}, validates the effectiveness of the fusion technique.

\textbf{Extended context-based generation:}
We notice that LLMs like GPT 3.5, PaLM, LLaMa, etc. seldom introduce random facts regarding the entities in the sentence, which are biased. For instance, we notice that some generations were unnecessarily sentimentally negative. Furthermore, for the rare entities, these LLMs tend to change the entity mentioned in the generated sentence, therefore rendering the generated sentence ambiguous and irrelevant. Therefore, in order to counter these issues, we introduce a new generation technique, where we scrape context for the entities from DBpedia articles. We then use PaLM to generate a single sentence connecting the entities using the relation with the help of the context. Through proper prompt instruction tuning, we ensured that we not get any extra information from the model's own knowledge, thus producing a less biased and entity-inclusive dataset.

\subsubsection{\textbf{Stage 3: Generated data evaluation using multiple quality parameters}}
\label{sec:qual_param}
Stage 3 deals with the first part of the second major question of comparing the quality of the sentences generated by various methods. We use traditional methods BLEU \citep{papineni2002bleu}, METEOR \citep{banerjee2005meteor} along with our own adaptations of recent works for \textit{grammar score}, \textit{sentiment score}, \textit{readability}, \textit{fluency}, \textit{accuracy}, \textit{relevance} and \textit{coherence}. The following sections provide details of the proposed adaptations to get different scores. 


\textbf{Grammar Score:}
The grammar score is determined using the following algorithm: 1) The \textbf{COEDIT} model by Grammarly \citep{raheja2023coedit} is employed to produce a grammatically accurate version of the input sentence. 2) Subsequently, the cosine similarity between the original sentence and the corrected sentence is calculated. 3) This cosine similarity value serves as the grammar score, ranging between 0 and 1, as rectifying the sentence grammatically does not alter its meaning. A higher score indicates the generation method's ability to generate grammatically correct sentences.

\textbf{Human Aligned Sentiment Score:}
Sentiment analysis is conducted using the base model \textbf{BERTweet}~\citep{pérez2023pysentimiento}. The model assigns sentiment scores of 0, 1, and 2, representing negative, neutral, and positive sentiments.  Our scoring method calculates the \textbf{absolute difference} between the sentiment scores of the generated sentence and its corresponding ground-truth `silver' sentence (see \ref{sec:temp_gen}), which avoids sentimental bias.  We then negate this difference, ensuring that higher final scores indicate better alignment with human sentiment. 

\textbf{TIGER Score:}
We use TIGERScore \citep{Jiang2023TIGERScoreTB} to compute errors in \textbf{fluency, accuracy, logical coherence, and relevance} of generated sentences. A default base error of 0 is given while computing errors, and when any error is flagged in any of these metrics, -1 is assigned to that metric corresponding to that sentence. So the higher the value, the better, as it represents fewer errors in the parameter.

\textbf{Dale-Chall Readability score:}
The Dale-Chall readability index~\citep{dale1948formula} is used to gauge the grade-level reading ability of any piece of text. Generally, a higher Dale-Chall readability score correlates to a more complex and difficult sentence. 


\subsubsection{\textbf{Stage 4: Ranker}}
    Stage 4 takes care of the second part of the second major question of ranking the different methods of sentence generation. Our objective is to pick the best sentences from the set of all the generated sentences for a particular tuple. In order to define the best generation, we need to rank the sentences leveraging the quality parameters discussed in section~\ref{sec:qual_param} also, as discussed in our ablation studies in \ref{sec: Stage 4 ranker ablation}. We propose a \textit{Ranker module}, inspired by \citep{jiang2023llmblender}, where we give weights to the quality parameters. The discussion about selecting weights given to each module can be found in appendix \ref{sec:ranker weights}. We compute the final \textbf{Sentence Evaluation Index (SEI)} using the weighted sum of normalized scores across all the chosen quality parameters. For every tuple, we compare each of the \textbf{11 generation techniques} (note that we do not include the extended context-based generation, gold and silver sentences in the ranker, the reason is discussed under the ablation studies \ref{sec: Stage 5 gold + ecb ablation}) based on the consolidated score SEI, and sort them to find the \textbf{top 3} generated sentences.

\subsubsection{\textbf{Stage 5: Blending the best}}
Once SEI has ranked the sentences, we amalgamate \citep{jiang2023llmblender} the top three generations along with the gold sentence and extended context-based generation using an LLM to output the final generation. The gold and extended context-based generation methods were chosen to be fused at the final round to reinforce less biasness, entity preservation, and conciseness. Furthermore, fusing the top three generations leads to improvement in the dataset diversity and quality. We perform the blending using the \textbf{Gemini-pro} model, with proper instruction tuning via promoting to act as a fuser and not to include any external knowledge other than the given five sentences for context. After fusion from Gemini, we get our final test dataset. The fusion prompt is discussed in appendix~\ref{sec:prompts}. We conducted an ablation study to highlight the utility of this stage in appendix \ref{sec: Stage 5 gold + ecb ablation}

\textbf{Consolidating final train, test, and validation sets: }
We generate sentences for 180k tuples by fusing generations from the top performing techniques i.e. LLaMa, GPT 3.5, and Flan T5, tuned on WebNLG with the corresponding gold and extended context-based generation method, leveraging Gemini-pro for fusion. From this set of generated sentences and tuples, we use 162k tuple-sentence pairs to form the train set and the remaining 18k to form the dev set. In a similar manner, we generate the final test set with 15k sentences and their corresponding tuples.  

\section{Results and Analysis}

This section compares our dataset's quality to other RE/RC datasets using benchmark models and LLM-based techniques (see Appendix \ref{sec: LLM based RC}) inspired by \citep{jiang2023llmblender, kasner-dusek-2020-data}. We also analyze LLM performance in terms of generation quality of the sentences along with sentiment, cost efficiency, and relation classification.
\label{sec:results}
\subsection{Dataset}
Table \ref{tab:dataset_statistics} summarizes the statistics of our dataset \texttt{AmalREC} along with other related datasets. AmalREC clearly contains a significantly higher number of relations with relatively better coverage for each relation type.
\input{Tables/dataset_statistics}
\input{Tables/Quality_Scores_reference_based}
\input{Tables/Quality_Scores_non_reference_based}
\input{Tables/all_baselines_results}

\subsection{Benchmark RC/RE models}
We evaluated state-of-the-art (SOTA) RC baselines, such as SpERT \cite{eberts2019span} and TANL \cite{simi-attardi-2016-adapting}, on our dataset, comparing it against established RE/RC datasets. SpERT struggles with relation classification on AmalREC as compared to ConLL04 and ADE. Our dataset poses a tougher challenge for relation classification, with SpERT showing a considerably low F1 score of \textbf{52.65\%}. Despite entity extraction being easier, relation classification proves more difficult on our dataset, making it a valuable addition to RC benchmark datasets.
Similarly, TANL, treating the task as a translation problem between augmented natural languages, faces challenges in achieving comparable performance on our dataset compared to others. The performance of AmalREC on relation extraction using TANL is worst than the existing datasets by having a relation extraction F1 score of \textbf{43.29\%}. AmalREC presents significant challenges for existing models in relation extraction and classification, indicating room for improvement in understanding diverse relations in natural language.

\subsection{Human Verification}
We extracted a uniformly shuffled random sample of 670 sentences from our dataset and verified it by human annotators. 2 reviewers were used per sentence. The reviewers classify every instance in the test set as correct, incorrect, or ambiguous. If we have contradictory opinions, then the sentence is termed ambiguous. If both agree on (in)correct, it is termed (in)correct. We observe 605 correct instances, implying \textbf{90.29\% accuracy}.

\section{Comparative Analysis of LLMs} 

The diverse LLMs employed in our study showcased varying proficiencies across a range of evaluation metrics, highlighting their unique strengths and limitations.

\subsection{Relation Classification} 

As evidenced in Table \ref{tab:RC_baselines}, GPT-3.5 consistently outperformed PaLM and Mixtral in relation classification tasks, achieving notably higher F1 scores in both zero-shot and few-shot settings. This suggests a superior ability to discern complex relationships within the AmalREC dataset, which features a significantly larger number of relation types compared to standard benchmarks like CoNLL04. PaLM and Mixtral, while still demonstrating reasonable performance, struggled to maintain comparable accuracy, particularly in the few-shot scenario.

\subsection{Sentence Quality and Sentiment}
Tables \ref{tab:reference_based} and \ref{tab:non_ref_based} evaluate sentence generation and sentiment alignment across various LLMs. Decoder-only models like GPT-3.5 and LLaMA achieved high readability scores, reflecting greater complexity, but often at the expense of accuracy compared to encoder-decoder models like Flan T5, which excelled in precision. Fusion models, combining the strengths of both types, balanced complexity and accuracy effectively. Despite most models generating slightly negative sentences compared to the 'silver standard', fusion models, especially those using GPT-3.5, showed improved Human-Aligned Sentiment (HAS) scores. This indicates that fusion models can better align with human sentiment and reduce biases. The comparative analysis in the appendix \ref{sec:average-sentence-length} reveals that LLaMA generates the longest sentences, while FlanT5(webNLG) produces the shortest. AmalREC, our final dataset, achieves a balanced approach with an average character count of 178.75 and an average word count of 26.55, integrating strengths from various methods and avoiding extremes in sentence length.  Additionally, LLMs tendency to produce negative sentiments was addressed using the HAS metric, with details provided in the appendix \ref{sec:error_gen}.

\subsection{Cost-Efficiency}
Table \ref{tab:cost_comparison} underscores the financial implications of LLM selection. GPT-3.5, while leading in performance, also incurred the highest costs. Mixtral, on the other hand, offered a more budget-friendly option, albeit with slightly lower accuracy. This trade-off between cost and performance necessitates careful consideration when deploying LLMs in real-world scenarios.

In summary, our comparison highlights the diverse capabilities of LLMs for relation extraction and classification. GPT-3.5 excels in accuracy, while fusion models balance complexity, accuracy, and sentiment alignment. The choice of LLM depends on task needs and budget, with future research aiming at optimization and cost-efficiency.
\section{Conclusion}
This work presents a pipeline framework for the generation of \textit{grammatically correct}, \textit{sentimentally human aligned}, \textit{readable}, \textit{fluent}, \textit{accurate}, \textit{relevant} and \textit{coherent} sentences from entity-relation tuples. As we leverage several generation techniques, including the recent LLMs, we propose a holistic evaluation method to rank them. We also provide a dataset of 204,399 sentences spanning 255 relational categories, a range broader than any of the existing datasets. We also evaluate the dataset on existing SOTA baselines for relation classification and extraction and show the need for further research in modeling for tasks on datasets with a many relation types.

\section{Future work}
Our work and proposed dataset can be used in many research avenues (a) \textbf{Relation Extraction and classification models}: As shown in the results~\ref{tab:RC_baselines}, the current SOTA models do not perform up to the mark when the number of relation types is high. So this suggests the need for further improvement in the modeling side. (b) \textbf{Complex ranking module}: we can further improve the ranking module by training a small ranking model on some human-annotated rankings of our dataset.  (c) \textbf{Using the taxonomical relational hierarchy \citep{parekh2019taxonomical}} for further relation-wise analysis of results. 


\section{Limitations}
\textbf{Premium LLMs}: Although we have evaluated the results on leading open-source and proprietary LLMs, our dataset has not been evaluated on LLMs like Claude-3 and GPT-4 owing to financial constraints. 

\noindent \textbf{Baselines comparison with other datasets}: LLM-based RC tasks could be explored on more datasets other than CoNLL04, which is not currently done owing to financial constraints.

\bibliography{custom}
\newpage
\appendix

\section{Appendix}
\label{sec:appendix}
\subsection{Weights for ranker}
\label{sec:ranker weights}
\begin{figure}[!ht]
\begin{center}
\includegraphics[width=0.5\textwidth]{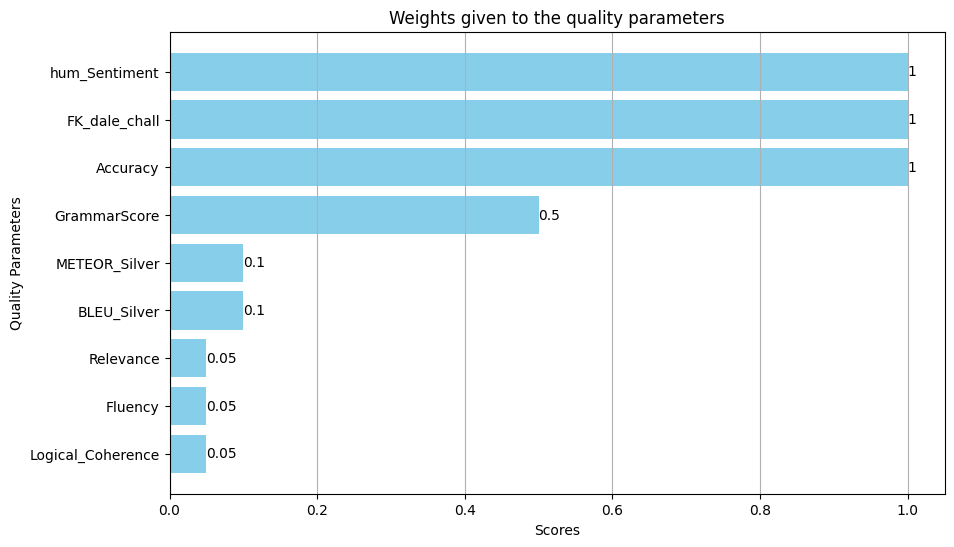}
\end{center}
\caption{Weights given to the quality parameters in the ranker module}
\label{fig:weights-SEI}
\end{figure}

Tables \ref{tab:reference_based} and \ref{tab:non_ref_based} present the quality scores used in the ranker. Silver sentences' Readability scores are low due to lack of complexity, and they are the highest for the combined dataset. Almost all the methods have given high scores for relevance, fluency, and logical coherence, indicating the low discriminatory power of these scores for ranking purposes. Hence, we assign low weights to relevance, fluency, and logical coherence in the ranking module. METEOR and BLEU scores are given slightly higher weights as they vary across generation techniques and contribute to generation quality.
Even though grammar scores don't vary much across generation techniques, they are important parameters, and hence, they are given a weight of 0.5. Finally, we give sentiment scores, readability, and accuracy the highest weights. Sentiment was given a high value to promote sentences that are sentimentally close to human sentiments. Readability was given a high score as it controls the complexity of the sentence, thus promoting high chances of diversity in the dataset. Finally, we want the sentences to be very accurate to ensure a relevant and unambiguous generation. After multiple experiments with the weights and human verification, we decided on a set of weights as shown in \ref{fig:weights-SEI}. Scores of the AmalREC indicate the quality of sentences as well as complexity. 

In the figure ~\ref{fig:weights-SEI} we have shown the weights assign to each of the scores. We have explained the rationale behind our choice of weights in the section ~\ref{sec:results}.

\subsection{Cost Analysis}
\label{sec:cost}
We have summarised our cost for running the baselines on the LLMs mixtral and GPT-3.5 in the table ~\ref{tab:cost_comparison}.
\input{Tables/cost_table}

\subsection{Average Sentence Length}
\label{sec:average-sentence-length}
\begin{figure}[t]
    \centering
    \includegraphics[width=0.5\textwidth]{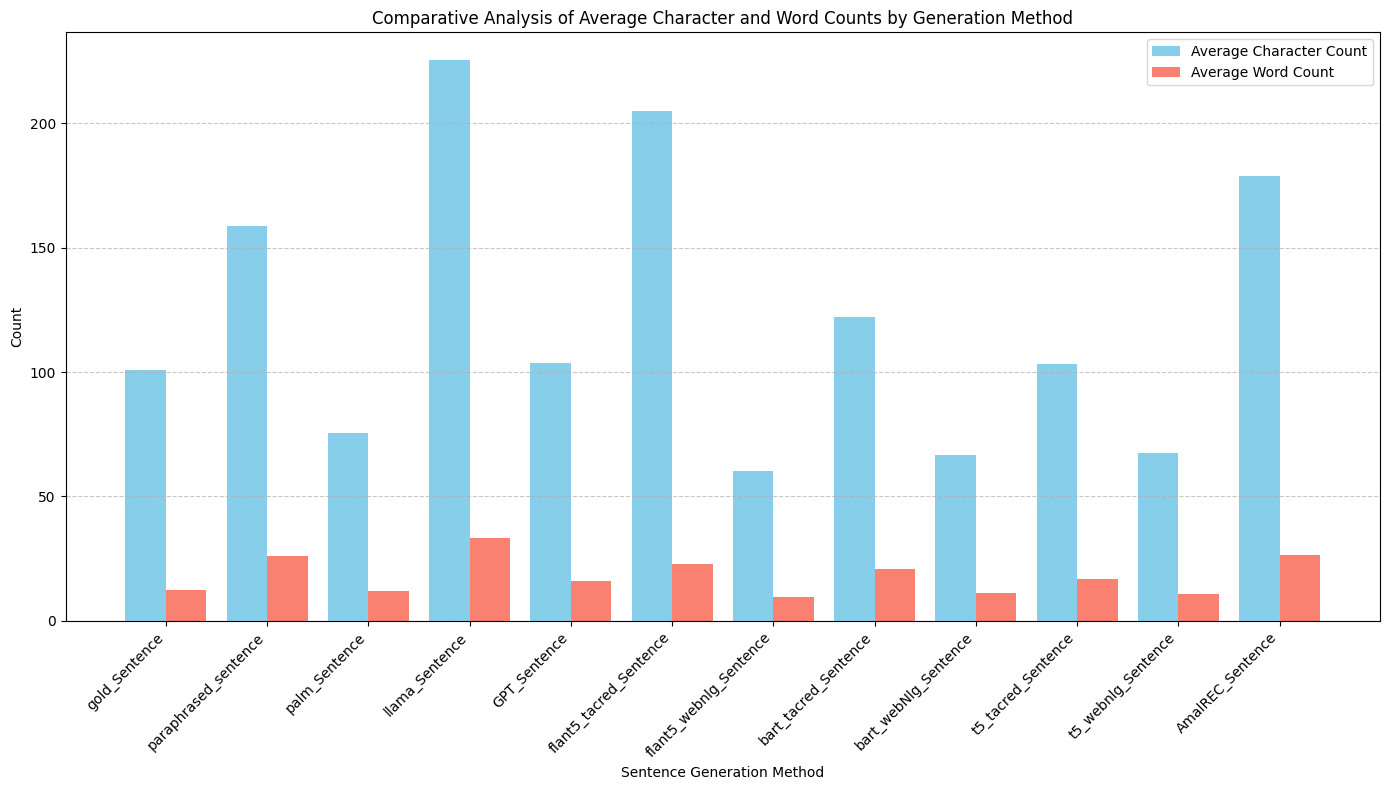}
    \caption{Average sentence length by all the generation techniques}
    \label{fig:avg_sentence_length}
\end{figure}
The average word and character counts of the various sentence creation techniques are found to differ significantly. The longest sentences are produced by LLaMA, which has an average word count of 33.38 and character count of 225.34. On the other hand, with an average word count of 9.69 and a character count of 60.32, FlanT5(webNLG) generates the shortest sentences. Higher averages in both metrics are shown by methods such as paraphrased(Silver) and FlanT5(TACRED), with average word counts of 22.78 and 158.74 and average character counts of 204.96, respectively as shown in figure \ref{fig:avg_sentence_length}. The findings demonstrate a notable degree of variation in sentence lengths among the various generating approaches, which can be attributed to their differing levels of complexity and density of material.

AmalREC, our final dataset, demonstrates a balanced approach by integrating the strengths and addressing the weaknesses of various techniques. With an average character count of 178.75 and an average word count of 26.55, the sentences in AmalREC avoids the extremes of both excessively long and overly brief sentences. It effectively merges the advantages of other methods, offering a more consistent and contextually appropriate sentence length. This fusion not only provides a more balanced length but also mitigates the issues of having excessively long sentences with extra-polated data or too short sentences, ensuring optimal performance and usability.

\subsection{Mapping module}
\label{sec: mapping module}
It is a filtration mechanism aimed at excluding such undesirable sentences during the merging process. In this mechanism, we integrate a \textbf{mapping module} into the prior generation stage to facilitate accurate alignment of sentences with the corresponding tuples. We have one-to-one mapping of tuples to the sentence, and therefore we performed an exact match of the entities in the respective sentences. The tuples which could not be mapped to any of the sentences were dropped and similarly the sentences which were not mapped to any of the tuples were dropped at the end to ensure a 100\% match in the final data. Further, in one relation bucket, the entity pairs are unique hence leaving no room for error in mapping in the final set. The prompts for LLM-based generation can be found in the appendix~\ref{sec:prompts}.

\subsection{LLM based RC}
\label{sec: LLM based RC}
Relation classification was performed using zero-shot and few-shot CoT prompting techniques on LLMs like PaLM, GPT3.5, and MixTral. Table \ref{tab:RC_baselines} summarizes the results. 
\textbf{Zero shot:}
To get an idea of LLMs' performance on the task with their pre-existing knowledge, we first test various LLMs using the Zero Shot prompting. Here, we present the model with the list of 255 relation tuples, a sentence, and two entities present in the sentence. We ask the model to classify the relation between the given entities in the sentence from the given list of tuples containing the entity types and the relation between them. The tuples for classification are of the form \textit{(Entity\_1\_Type, Relation, Entity\_2\_Type)}. \textbf{Few shot COT:}
In Few shot Chain Of Thought prompting \citep{wei2022chain}, We provide the model with a few examples of the relation classification from the given tuples along with a reasoning for the classification. We also provide the list of 255 tuples, a sentence, and two entities present in the sentence as done in the Zero-shot prompting. The prompts for both zero-shot and few-shot are provided in the appendix ~\ref{sec:prompts}. GPT3.5 has the best results for our dataset among the 3 LLMs. In comparison with CoNLL04, our dataset is more difficult to solve.

\subsection{Baseline model's Configurations}
\label{sec:config}
The Eval configuration of SpERT model below:
\begin{verbatim}
rel_filter_threshold = 0.4
size_embedding = 25
prop_drop = 0.1
max_span_size = 10
store_predictions = true
store_examples = true
sampling_processes = 4
max_pairs = 1000
tokenizer=bert-base-cased
\end{verbatim}
The Train configuration and hyper parameters for SpERT are as below:
\begin{verbatim}
neg_entity_count = 10
neg_relation_count = 10
epochs = 20
lr = 5e-5
lr_warmup = 0.1
weight_decay = 0.01
max_grad_norm = 1.0
rel_filter_threshold = 0.4
size_embedding = 25
prop_drop = 0.1
max_span_size = 10
max_pairs = 1000
tokenizer=bert-base-cased
\end{verbatim}

The eval, train and test setting for TANL for relation classification are:
\begin{verbatim}
multitask = True
model_name_or_path = t5-base
num_train_epochs = 10
max_seq_length = 512
train_split = train
per_device_train_batch_size = 4
do_train = True
do_eval = True
do_predict = True
chunk_size = 512
chunk_overlap = 96
chunk_size_eval = 512
chunk_overlap_eval = 96
\end{verbatim}

The eval, train and test setting for TANL for relation extraction are:
\begin{verbatim}
model_name_or_path = t5-base
num_train_epochs = 10
max_seq_length = 512
max_seq_length_eval = 512
per_device_train_batch_size = 4
per_device_eval_batch_size = 4
do_train = True
do_eval = True
do_predict = True
chunk_size = 512
chunk_overlap = 96
chunk_size_eval = 512
chunk_overlap_eval = 96
\end{verbatim}

\subsection{Fine-tuning hyper-parameters}
\label{sec:finetuning hp}
The following parameters were specified during model training. The num train epochs is set to 7, this indicated the total number of times the entire training dataset will be iterated over during the training process.
The batch size for training and evaluation on each device are controlled by per device train batch size and per device eval batch size, with a size of 2 for both. The warmup steps and weight decay parameters contribute to the optimization process by adjusting the learning rate schedule
and controlling the amount of regularization, respectively.
Logging of training information is configured with logging dir, specifying the directory for logs, and logging steps indicating the frequency at which training logs will be recorded (every 100 steps). Additionally, the evaluation strategy is set to ’steps’, and evaluation is performed every 100 steps (eval steps).
Model saving is managed with save total limit (limiting the total number of saved check- points to 2) and save steps (saving a checkpoint every 500 steps). The learning rate is set to 2e-5, controlling the step size during optimization. Lastly, gradient accumulation steps is configured at 8, influencing how many forward and backward passes are performed before a parameter update.

\subsection{Model-based prompt samples}
\label{sec:prompts}
\textbf{Relation Classification: Prompt for Zero-shot prompting}

Your task is to classify the relationship between two entities mentioned in a sentence.
The entities will be provided to you along with the sentence.
Your goal is to identify the relation between those 2 entities and the type of the 2 entities to find the unique tuple: (Entity\_1\_type, Relation, Entity\_2\_type)
Classify it into one of the following buckets: $\{$classes$\}$
Classify the following sentence into one of the above relation tuples, and only output the class in json format which should be from one of the above classes:

Sentence: \{s\} \\
Entity1: \{e1\} \\
Entity2: \{e2\} \\
Label:

\{ \\
    "Entity\_1\_type": \\
    "Relation": \\
    "Entity\_2\_type": \\
\}


\textbf{Prompt for Few-Shot CoT prompting}Your task is to classify the relationship between two entities mentioned in a sentence.
The entities will be provided to you along with the sentence.
Your goal is to identify the relation between those 2 entities and the type of the 2 entities to find the unique tuple: (Entity\_1\_type, Relation, Entity\_2\_type)
Classify it into one of the following buckets:
\{classes\}

Below is the example of the task:

Sentence: Hamburg is the administrative district of Blankenese. \\
Entity1: Blankenese \\
Entity2: Hamburg \\
Label:

\{ \\
    "Reasoning": "Here Blankenese as well as Hamburg is a location and Hamburg is administrative district of Blankenese so the relation between the entity is administrativeDistrict.",\\
    "Entity\_1\_type": "Location",\\
    "Relation": "administrativeDistrict",\\
    "Entity\_2\_type": "Location"\\
\}

\{More examples\}

Classify the following sentence into one of the above relation tuples, and only output the class in json format which should be from one of the above classes. Also provide thorough reasoning in the reason key for your classification.:

Sentence: \{s\} \\
Entity1: \{e1\} \\
Entity2: \{e2\} \\
Label:
\{
    "Reasoning": \\
    "Entity\_1\_type": \\
    "Relation": \\
    "Entity\_2\_type":
\}

The \{classes\}, \{s\}, \{e1\}, \{e2\} are variables containing the list of tuples, the sentence for relation classification, the entity1 and entity2 respectively.
The \{More examples\} variable contains the list of examples.

\textbf{Prompt used or generation after fine-tuning Flan T5}
Generate a sentence, given the following entities and reaction between them: 
Entity 1:e1
Entity 2: e2
Relation: rel

\textbf{Prompt for amalgamation}
Given the entity, entity type and the relation between the entity as a tuple , Generate a diverse sentence with using the given entities and the relation between them.
For supporting knowledge and adding diversity we are providing some sentences already generated using this tuple.
So use them for diverse sentence generation.
Do not add any extra information apart from the given 5 sentences.
Make sure the relation between the entities is preserved while keeping the sentence tough for relation classification task.
Make sure to use information in the all 5 sentences.\\
Tuple\_format: (E1, E1\_TYPE, RELATION, E2, E2\_TYPE)\\
Tuple: \{r\_tuple\} \\
Sentence1: \{s1\} \\
Sentence2: \{s2\} \\
Sentence3: \{s3\} \\
Sentence4: \{gold\} \\
Sentence5: \{ECB\} \\

Now create a diverse, difficult sentence using the above information.
Make sure to avoid using any extra information apart from the given 5 sentences.

Here variable \{s1\}, \{s2\}, \{s3\} are the top 3 sentences from the ranker, \{gold\} is the gold sentence and \{ECB\} is the extended context based sentence.

\subsection{Erroneous generations}
\label{sec:error_gen}
Please refer to \ref{table:errors} and \ref{table:errors_method} for details on the types of errors encountered across various generation methods. We have categorized these errors based on the generation method, tuple, and sentence. Notably, Flan T5's TACRED model exhibited a tendency to produce negative sentiments and propagate hate speech, even in the absence of such data in the input tuples. For instance, it inappropriately associated the term "Kerala" with terrorism. Similar issues were observed with the BART TACRED model. Our analysis suggests that the TACRED dataset, which includes data from news articles, may contribute to such biases. Additionally, some relations were not clearly defined by the LLMs, leading to inconsistencies, and there were instances of repetition in the outputs of smaller LLMs. Hallucinations are a common drawback of LLMs, and we observed this issue in our study, resulting in the generation of factually incorrect data. In Table \ref{table:errors_method}
, we can observe how Flan T5 exhibited specific errors in few-shot and zero-shot settings without fine-tuning. The model generated irrelevant data and repeated small, out-of-context sentences based on the provided tuples. Although the prompt specified not to mention entity types in the sentences, it was observed that this guideline was not followed by the LLM.

\input{Tables/erroroneous_generations_example}
\input{Tables/issues_with_few_shot}

\subsection{Ablation Studies}

\subsubsection{Why was finetuning required in Stage 2?}
\label{sec: Stage 2 finetuning ablation}
\begin{figure}[ht]
\begin{center}
\includegraphics[width=0.5\textwidth]{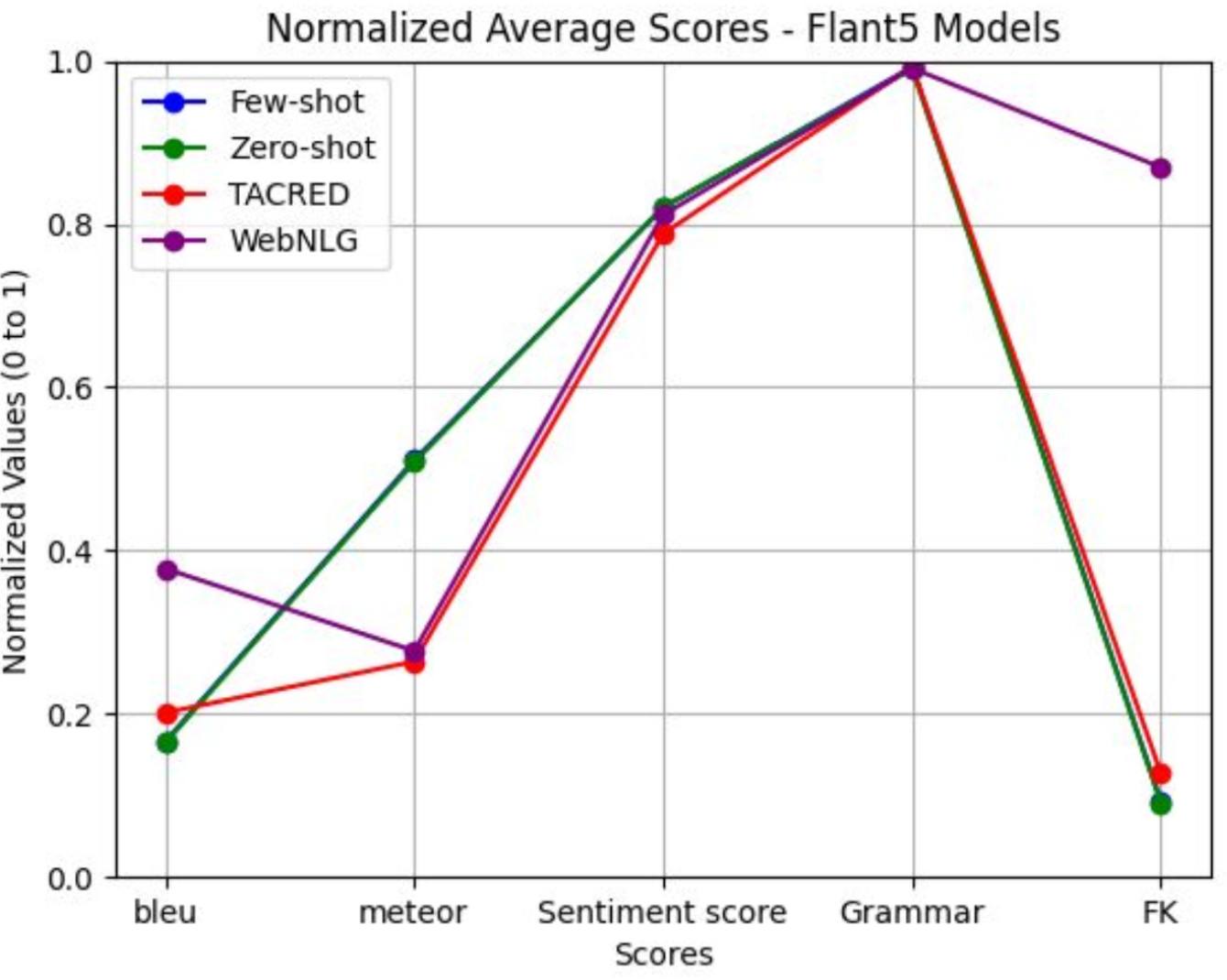}
\end{center}
\caption{FlanT5 results on zeroshot, fewshot and finetuning, demonstrating the need for finetuning}
\label{fig:ablation-finetuning}
\end{figure}

We finetuned encoder-decoder models instead of using zeroshot or fewshot prompting for them. This section presents an ablation study on finetuning, thus comparing the results of zeroshot and fewshot prompts against instruction finetuning on models. In Figure \ref{fig:ablation-finetuning}, we show the example of FlanT5, and use a few baseline metrics to compare the generations. Finetun- ing is performed on WebNLG and TACRED. Finetuning on WebNLG proved to drastically improve BLEU scores, and also helped in improving Human aligned sentiment scores, Grammatical accuracy and readability difficulty levels. Zeroshot produces the simplest sentences, thus showing a higher METEOR score. As discussed in Table \ref{table:errors_method}, Flan T5's performance was significantly affected, with common errors degrading sentence quality. This underscores the necessity for fine-tuning to improve its performance.

\subsubsection{Why was fusion methodology required in Stage 2?}
\label{sec: Stage 2 fusion ablation}
\begin{figure}[!ht]
\begin{center}
\includegraphics[width=0.5\textwidth]{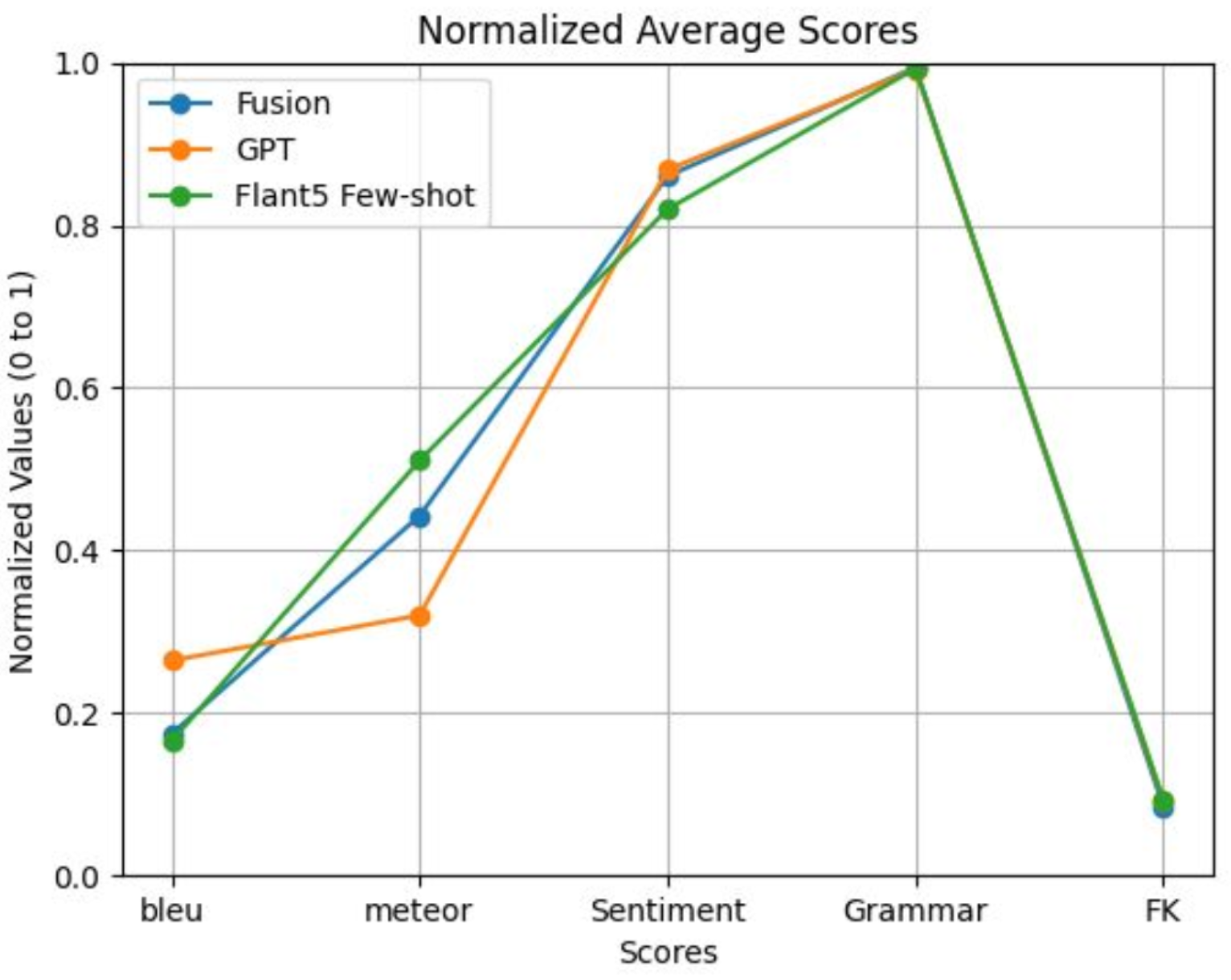}
\end{center}
\caption{Fusion technique results against decoder only and encoder-decoder models separately}
\label{fig:ablation-fusion}
\end{figure}

An ablation study was performed to determine the necessity of inclusion of the fusion technique as a method of generation. Observe Fig. \ref{fig:ablation-fusion}, The fusion technique helps to improve the HAS scores from the scores on generations by encoder-decoder models that weren’t finetuned on data generated from decoder-only models. The METEOR scores show that fusion models help to keep the sentences simpler than the complex generations by LLaMa or GPT while also making them more complex than the simple FlanT5 or T5 generations.

\subsubsection{What is the benefit of the ranker module, Stage 4?}
\label{sec: Stage 4 ranker ablation}
The ranker module leverages picking up the sentences of desired quality as per the weights given to the quality parameters. It facilitates quality generation by giving only those sen- tences which meet the decided quality standards set at the best, to the fuser at the 5th stage, thus facilitating quality generation. To test this hypothesis, we conduct this ablation study where we remove the ranker module and directly fuse all the generation techniques. For the purpose to testing, we pick a subset of the test set with more than 500 data points, spanning across all the 255 relation types. We can see the quality of the generated sentences when we remove the ranker module, and its comparison with the other generation techniques in Table \ref{tab:ablation_ranker}. From the table, we can easily state that ranker based generation technique gave better results on almost all the quality parameters.

\input{Tables/Ablation_Ranker_Justification}

\subsubsection{Why are Gold and ECB added for the last round in Stage 5?}
\label{sec: Stage 5 gold + ecb ablation}
We propose the following hypothesis as to why these generations are important: ECB or extended context based generations are guaranteed to be factually correct and the facts are updated as they are scraped from the web-pages directly, Gold generations are generated by humans and hence they are grammatically sound and structurally exemplar for our dataset. These generations thus must be included in our blending stage, bypassing the need for them to pass stage 4. The top 3 generations from stage 4 play the roll of adding diversity to our generations.
We conduct this ablation study to bolster our above hypothesis. We have thus compared the scores on generations obtained by blending the top 3 generations of the ranker against the scores obtained from blending the top 3 generations from the ranker along with gold and ECB. This is shown in Table \ref{tab:Ablation_stage5}. From the analysis presented in the table, we can comment that introducing the Gold and the ECB sentence in the last step improved the quality of AmalREC on almost all the quality parameters.

\subsection{Reproducibility statement}
We are committed to fostering reproducibility within our research. To this end, we will make our code publicly available through a GitHub upon publication. We primarily utilize publicly available datasets and adhere to all ethical guidelines and usage restrictions specified by the original authors.  We provide detailed descriptions of all models utilized, including architectures, hyperparameters, costs, and computing resources in the appendix section ~\ref{sec:cost}, ~\ref{sec:config}, ~\ref{sec:finetuning hp}. Prompting methods, annotations, and dataset splits are clearly documented. While we leveraged AI-based tools for grammar correction and coding assistance, we emphasize that the core concepts, research directions, and experimental execution were entirely independent of AI involvement. We acknowledge the inherent variability associated with large language models and disclose any measures taken for reproducibility (e.g., fixed temperature settings). By providing these resources, we empower the research community to thoroughly reproduce our results, validate our findings, and build upon our work. 

\subsection{Ethics statement}
We affirm that our work adheres to all ethical standards in research and publication. We have adhered to all the standards set by the LLMs and proprietary LLMs and platforms used in our work. Various ethical considerations have been carefully addressed to ensure the responsible and equitable application of computational linguistics methodologies. Detailed information is provided to facilitate the reproducibility of our results. Transparency is maintained through the sharing of code, datasets, and other pertinent resources, thereby enabling the research community to validate and further build upon our findings. The claims presented in the paper are consistent with the results of our experimentation. However, it is acknowledged that a certain degree of deviation is inherent in proprietary large language models. Our work is meticulously described, including annotations, dataset splits, models utilized, hyperparameter settings, quality parameters calculation and trends explanation, prompts for the generation techniques and prompting methods experimented with, thereby ensuring the reproducibility of our research.

\input{Tables/Ablation_stage5}

\end{document}

%% file: Tables/dataset_statistics.tex
\begin{table}[!tb]
    \centering
    \small 
    \begin{tabular}{lrrrrr}\toprule
        \textbf{Dataset} & \textbf{\#Rel} & \textbf{\#Train} & \textbf{\#Dev} & \textbf{\#Test} & \textbf{\#Total} \\\midrule
        SemEval & 9 & 6,507 & 1,493 & 2,717 & 10,717\\ 
        TACRED & 41 & 68,124 & 22,631 & 15,509 & 106,264\\ 
        SciERC & 7 & 16,872 & 2,033 & 4,088 & 22,993\\ 
        CoNLL & 5 & 922 & 231 & 288 & 1,441\\ 
        ACE05 & 6 & 121,368 & 27,597 & 24,420 & 173,385\\ 
        \textbf{AmalREC} & \textbf{255} & \textbf{170,416} & 18,936 & 15,047 & \textbf{204,399}\\ \bottomrule
    \end{tabular}
    \caption{Comparison of datasets in terms of relation types and subset sizes}
    \label{tab:dataset_statistics}
\end{table}

%% file: Tables/Quality_Scores_reference_based.tex
\begin{table*}[!tb]
    \centering
    \small
    \begin{tabular}{llccc}
    \toprule
        &\textbf{Generation methods} & \textbf{BLEU} & \textbf{METEOR} & \textbf{HAS} \\\midrule
        ~ & Flan T5\_TACRED & 0.2273 & 0.5056 & -0.3694 \\ 
        ~ & Flan T5\_webNLG & 0.2191 & 0.5651 & -0.3575 \\ 
        Fine tuning LLM & BART\_TACRED & 0.2402 & 0.3116 & -0.4655 \\ 
        ~ & BART\_webNLG & 0.2258 & 0.5587 & -0.3558 \\ 
        ~ & T5\_TACRED & 0.2324 & 0.3540 & 0.4112 \\ 
        ~ & T5\_webNLG & 0.2128 & 0.5387 & 0.3561 \\ \midrule
        ~ & LLaMA & 0.2441 & 0.3129 & -0.4665 \\ 
        Few shot LLM & GPT & 0.2397 & 0.4771 & -0.2731 \\ 
        ~ & PaLM & 0.2272 & 0.5135 & -0.3534 \\ \midrule
        ~ & Fusion\_Flan T5\_LLaMA & 0.2461 & 0.3693 & -0.4388 \\ 
        Fusion & Fusion\_Flan T5\_GPT & 0.2259 & 0.5662 & -0.3670 \\ 
        ~ & Fusion\_Flan T5\_PaLM & 0.2214 & 0.5767 & -0.3573 \\\midrule
        \textbf{AmalREC}  & & \textbf{0.2532} & 0.2915 & -0.2834 \\\bottomrule
    \end{tabular}
    \caption{BLEU, METEOR and Human-Aligned Sentiment Scores with silver as a reference.}
    \label{tab:reference_based}
\end{table*}

%% file: Tables/Quality_Scores_non_reference_based.tex
\begin{table*}[!tb]
    \centering
    \small
    \resizebox{\textwidth}{!}{
    \begin{tabular}{lllllll}
        \toprule
        \textbf{Generation Method} & \textbf{Fluency} & \textbf{Accuracy} & \textbf{Coherence} & \textbf{Relevance} & \textbf{Grammar} & \textbf{Readability} \\
        \midrule
        Flan T5\_TACRED          & -0.0989    & -0.4739    & -0.0265    & -0.0989    & 0.9934    & 11.3738 \\
        Flan T5\_webNLG          & -0.0581    & -0.3094    & -0.0088    & -0.0581    & 0.9910    & 11.9293 \\
        BART\_TACRED            & 0          & -0.1951    & 0          & 0          & 0.9938    & 10.2455 \\
        BART\_webNLG            & 0          & -0.3111    & 0          & 0          & 0.9908    & 11.6185 \\
        T5\_TACRED              & 0          & -0.1830    & 0          & 0          & 0.9936    & 10.2216 \\
        T5\_webNLG              & 0          & -0.2367    & 0          & 0          & 0.9680    & 12.1631 \\\midrule
        LLaMA                   & 0          & -0.1452    & 0          & 0          & 0.9972    & 12.3532 \\
        GPT-3.5                 & 0          & -0.2254    & 0          & 0          & 0.9916    & 12.2486 \\
        PaLM                    & 0          & -0.3375    & 0          & 0          & 0.9774    & 12.0627 \\\midrule
        Fusion\_Flan T5\_LLaMA   & 0          & -0.2039    & 0          & 0          & 0.9984    & 11.7320 \\
        Fusion\_Flan T5\_GPT     & 0          & -0.3217    & 0          & 0          & 0.9918    & 12.0415 \\
        Fusion\_Flan T5\_PaLM    & 0          & -0.3361    & 0          & 0          & 0.9894    & 12.1635 \\
        Silver                  & 0          & -0.2519    & 0          & 0          & 0.9890    & 10.7670 \\\midrule
        
        \textbf{AmalREC} & \textbf{-0.0007} & \textbf{-0.1502} & 0 & 0 & 0.9968 & \textbf{12.5877} \\ 
        \bottomrule
    \end{tabular}
    }
    \caption{Scores on Non-reference based evaluation metrics indicating the negation of error in accuracy, fluency, coherence and relevance, grammatical correctness and complex readability}
    \label{tab:non_ref_based}
\end{table*}

%% file: Tables/all_baselines_results.tex
\begin{table*}[!tb]
    \centering
    \small
    \resizebox{\textwidth}{!}{ 
    \begin{tabular}{l|l|lllllll}
    \toprule
        ~ & ~ & ~ & ~ & ~ & RC & ~ & ~ & ~ \\ 
        \midrule
        Model & Dataset & F1 & ~ & Precision & ~ & Recall & ~ & Accuracy \\ 
        \midrule
        PaLM & CoNLL04(Zero Shot) & 84.41 & ~ & 94.49 & ~ & 79.62 & ~ & 79.62 \\ 
        ~ & CoNLL04(Few Shot) & 87.62& ~  & 95.61 & ~ & 81.86 & ~ & 81.27 \\ 
        ~ & AmalREC(Zero shot) & 43.94 & ~ & 69.66 & ~ & 39.71 & ~ & 39.6 \\ 
        ~ & AmalREC(Few shot) & 35.33 & ~ & 57.14 & ~ & 34.407 & ~ & 27.97 \\ 
        \midrule
        Mixtral & CoNLL04(Zero Shot) & 61.74 & ~ & 91.68 & ~ & 59.61 & ~ & 59.63 \\ 
        ~ & CoNLL04(Few Shot) & 85.13 & ~ & 92.66 & ~ & 80.58 & ~ & 77.90 \\ 
        ~ & AmalREC(Zero shot) & 31.13 & ~ & 29.64 & ~ & 29.64 & ~ & 29.63 \\ 
        ~ & AmalREC(Few shot) & 37.72 & ~ & 35.61 & ~ & 35.61 & ~ & 35.5 \\  
        \midrule
        GPT3.5 & CoNLL04(Zero Shot) & 87.2 & ~ & 91.51 & ~ & 86.49 & ~ & 86.49 \\ 
        ~ & CoNLL04(Few Shot) & 79.78 & ~ & 89.03 & ~ & 76.0 & ~ & 75.83 \\ 
        ~ & AmalREC(Zero shot) & 52.56 & ~ & 50.49 & ~ & 50.49 & ~ & 50.49 \\ 
        ~ & AmalREC(Few shot) & 48.66 & ~ & 44.655 & ~ & 44.655 & ~ & 44.64 \\ 
        \midrule 
        ~ & ~ & E(F1) & R(F1) & E(Precision) & R(Precision) & E(Recall) & R(Recall) & Accuracy \\\midrule
        SpERT & CoNLL04 & 88.94 & 71.47 & 88.25 & 73.04 & 89.64 & 70.00 \\ 
        ~ & ADE & 88.95 & 79.24 & 88.69 & 78.09 & 89.2 & 80.43 \\ 
        ~ & SciERC & 67.62 & 46.44 & 68.53 & 49.79 & 66.73 & 43.53 \\ 
        ~ & AmalREC & 90.58 & 52.65 & 90.43 & 48.22 & 91.07 & 57.01 \\ 
        \midrule
        TANL & TACRED & - & 71.90 & - & ~ & - & ~ & - \\ 
        ~ & FewRel 1.0 & - & 82.60 & - & ~ & - & ~ & - \\ 
        ~ & AmalREC & - & 79.19 & - & 84.90 & - & 75.27 & 75.37\\ 
        \midrule
        ~ & ~ & ~ & ~ & ~ & RE & ~ & ~ & ~ \\ 
        \midrule
        TANL & CoNLL04 & 89.40 & 71.40 & - & ~ & - & ~ & - \\ 
        ~ & ACE2005 & 88.9 & 63.7 & - & ~ & - & ~ & - \\
        ~ & ADE & 90.2 & 80.6 & - & ~ & - & ~ & - \\
        ~ & NYT & 94.9 & 90.8 & - & ~ & - & ~ & - \\
        ~ & AmalREC & 77.17 & 43.29 & 91.74 & 61.37 & 66.60 & 33.44 & 33.94 \\
        \bottomrule
    \end{tabular}
    }
    \caption{Comparison across different RC datasets on SOTA baselines. Table abbreviations: E(F1) and R(F1) represent the F1 score for entity extraction and F1 score for RE respectively.}
    \label{tab:RC_baselines}
\end{table*}

%% file: Tables/cost_table.tex
\begin{table}[!ht]
\centering
\begin{tabular}{@{}llr@{}}
\toprule
Model & Method & Cost (in \$) \\ \midrule
Mixtral & Zero shot & 8 \\
 & Few Shot COT & 12 \\
GPT 3.5 & Zero shot & 13 \\
 & Few Shot COT & 17 \\ \bottomrule
\end{tabular}
\caption{Costs for Different Models and Methods}
\label{tab:cost_comparison}
\end{table}

%% file: Tables/erroroneous_generations_example.tex
\begin{table*}[ht]
\centering
\begin{tabular}{|p{2.5cm}|p{3cm}|p{5cm}|p{4.7cm}|}
\hline
\textbf{Type of Error} & \textbf{Paradigm} & \textbf{Tuple} & \textbf{Sentence} \\ 
\hline
Hallucinations & Flan T5-Tacred & Jody Carr, Person, predecessor, Roland Haché, Person & Jody Carr was the daughter of a man named Roland Haché and she was born in the year 1922. \\ 
\hline
Controversial Generation & T5 Tacred & Kerala\_Nadvatul\_Mujahideen, Organization, locatedAt, Kerala, Location & The Kerala Nadvatul Mujahideen (KNM) is a terrorist organization based in Kerala. \\ 
\hline
Missing the relation & 1. FlanT5 WebNLG 2. FlanT5 Tacred & Death\_of\_Elli\_Perkins, Person, religion, Scientology, Organization & 
1. The death of Elli Perkins was a result of Scientology. \newline
2. The death of Elli Perkins was announced by the Scientologists. \\ 
\hline
Inconsistency when entity overlap & FlanT5 Tacred & Kendriya Vidyalaya Raebareli, authority, Kendriya Vidyalaya, Organization & The Kendriya Vidyalaya Raebareli is an organization. \\ 
\hline
Unsolicited Accusation & BART Tacred & Rajnish Kumar (politician), Person, residence, Mukerian, Location & Rajnish Kumar, a member of the opposition and a former cabinet minister, is accused of protecting the gang-rape of a woman at his residence in the holy city of Mukerian, which is south of Delhi. \\ 
\hline
Missing entities & BART T5 & Lorella Stefanelli, Person, associatedWith, Nicola Renzi, Person & The leader of the Democratic People's Republic of Italy, Matteo Renzi, met with his sister-in-law, former First Lady Lorella Stefanelli, in the northern Italian city of Perugia, on Thursday to discuss the situation. \\ 
\hline
Repetitions & FlanT5 Tacred & Craig Foss, Person, predecessor, Rick Barker, Person & Rick Barker was the person who was the person who was the person who was the person who.. \\ 
\hline
Irrelevant data from prompt & T5 Tacred & Nanga Linsum, Location, country, Malaysia, Location, Nanga Linsum & The tuple information is a tuple that is used to describe the location of the embassy in Malaysia, which is located in the city of Kuala Lumpur. \\ 
\hline
\end{tabular}
\caption{Error types and examples from different paradigms.}
\label{table:errors}
\end{table*}

%% file: Tables/issues_with_few_shot.tex
\begin{table*}[ht]
\centering
\begin{tabular}{|p{3cm}|p{2.5cm}|p{5cm}|p{5cm}|}
\hline
\textbf{Type of Error} & \textbf{Method} & \textbf{Input triple} & \textbf{Sentence} \\ 
\hline
Mentioning Entity types in sentences & Few-shot & National African Religion Congress, Organization, locatedAt, United States, Location & The National African Religion Congress is located in the United States. It is an organization. \\ 
\hline
Spelling mistakes/Assumptions & Zero-shot & Christ Church, Childers, Location, locatedAt, Queensland, Location & Christ Church, Childers is located in Queensland... Queensland.. Queensland.. \\ 
\hline
Not able to justify relation between entities & 1. Few-shot 2. Zero-shot & 1. KirstieAlley, Person, religion, Scientology, Organization \newline 2. Jenna Syken, Person, formerChoreographer, Anjelika Krylova, Person & 1. Kirstie Alley is a person and Scientology is an organization. \newline 2. Jenna Syken is a former choreographer. Anjelika Krylova is a person. \\ 
\hline
Contradiction / Incorrect data & Zero-shot & Datia, Location, country, India, Location & Datia is a country in India. It is a location. \\ 
\hline
\end{tabular}
\caption{Examples of generations from FlanT5 few shot and zero shot, without finetuning.}
\label{table:errors_method}
\end{table*}

%% file: Tables/Ablation_Ranker_Justification.tex
\begin{table*}[h!]
\centering
\small
\begin{tabular}{lcc}
\hline
\textbf{Scores} & \textbf{AmalREC} & \textbf{Generations without Ranker} \\ \hline
METEOR\_Gold & 0.317485 & 0.300968 \\ 
METEOR\_Silver & 0.296490 & 0.278150 \\ 
Sentiment & 1.519685 & 1.653543 \\
FK\_Reading\_Ease & 18.588175 & 16.002220 \\
GrammarScore & 0.497408 & 0.498958 \\
TigerScore\_Relevance & 0.000000 & 0.000000 \\
TigerScore\_Fluency & 0.000000 & 0.000000 \\
TigerScore\_Accuracy & -0.135827 & -0.155512 \\
TigerScore\_Logical\_Coherence & 0.000000 & 0.000000 \\ \hline
\end{tabular}
\caption{Comparison of AmalREC and Generations without Ranker across different scores.}
\label{tab:ablation_ranker}
\end{table*}

%% file: Tables/Ablation_stage5.tex
\begin{table*}[h!]
\centering
\small
\begin{tabular}{lcc}
\hline
\textbf{Scores} & \textbf{AmalREC} & \textbf{Generation without Gold and ECB} \\ \hline
BLEU            & 0.265434          & 0.263208                                 \\ 
Sentiment       & 1.522954          & 1.530938                                 \\ 
FK Grade Level  & 17.161685         & 16.114491                                \\ 
FK Dale Chall   & 12.601888         & 12.471861                                \\ 
GrammarScore    & 0.497386          & 0.498329                                 \\ 
TigerScore Relevance & 0.000000     & 0.000000                                 \\ 
TigerScore Fluency   & 0.000000     & 0.000000                                 \\ 
TigerScore Accuracy  & -0.131737    & -0.165669                                \\ 
TigerScore Logical Coherence & 0.000000 & 0.000000                            \\ \hline
\end{tabular}
\caption{Comparison of AmalREC and Generation without Gold and ECB on various scores.}
\label{tab:Ablation_stage5}
\end{table*}